\documentclass[prb,superscriptaddress,twocolumn]{revtex4-1}
\usepackage{lineno}
\usepackage{epsfig}

\usepackage{graphicx}
\usepackage{dcolumn}
\usepackage{bm}
\usepackage{tabularx}
\usepackage{hyperref}
\usepackage{lipsum}
\usepackage{tablefootnote}
\usepackage{multirow}
\usepackage{ulem}
\usepackage{comment}
\usepackage{booktabs}

\usepackage{subfig}
\hypersetup{colorlinks=true, citecolor=blue, filecolor=blue, linkcolor=blue, urlcolor=blue}
\newcolumntype{Y}{>{\centering\arraybackslash}X}
\begin{document}
\title{PyLRO: A Python Calculator for Analyzing Long-Range Structural Order}

\author{Kevin Parrish}
\email{parrik1@unlv.nevada.edu}
\affiliation{Department of Physics and Astronomy, University of Nevada, Las Vegas, NV 89154, USA}

\author{Qingyang Hu}
\email{qingyang.hu@hpstar.ac.cn}
\affiliation{Center for High Pressure Science and Technology Advanced Research, Beijing 100193, China}

\author{Qiang Zhu}
\email{qzhu8@charlotte.edu}
\affiliation{Department of Mechanical Engineering and Engineering Science, University of North Carolina at Charlotte, Charlotte, NC 28223, USA}

\date{\today}

\begin{abstract}
We present \texttt{PyLRO}, an open-source Python calculator designed to detect, quantify, and display long-range order in periodic structures. The program’s design methodology, workflow, and approach to order quantification are described and demonstrated using a simple toy model. Additionally, we apply \texttt{PyLRO} to a series of metastable AlPO$_4$ structural intermediates from a prior high-pressure study, demonstrating how to compute and visualize structural order in all directions on a Miller sphere. We further highlight the program’s capabilities through a high-throughput analysis of structural patterns in the pressure-induced amorphization of AlPO4, revealing atomistic insights within specific energy regions of massive amorphous structures. These results suggest that \texttt{PyLRO} can be a valuable tool for investigating crystal-amorphous transition in materials research.
\end{abstract}

\vskip 300 pt
\maketitle
\section{Introduction}

Broadly speaking, materials can be classified as crystalline, quasi-crystalline, and noncrystalline based on their atomic ordering. The long-range translational symmetries define the structure of crystalline materials and give rise to many fascinating properties for materials science \cite{Evarestov2007}. Much of crystallographic theory is built upon the concept of idealized, perfectly ordered crystals, which provides a framework for understanding symmetry-breaking phenomena. In contrast, noncrystalline solids, often referred to as amorphous solids, lack these long-range symmetries, leading to a much wider variety of material types. The challenge of identifying useful materials from the vast space of possible amorphous solids is formidable. Crystallographic theories, such as Zachariasen’s random network theory for glass formation \cite{doi:10.1021/ja01349a006}, have been fairly successful in explaining amorphous solid formation by applying crystal chemistry rules and patterns. However, much remains to be explored, particularly concerning the extent to which some amorphous solids still conform to crystallographic principles of long-range order.

A deeper structural connection between crystalline and noncrystalline solids can be understood by examining the formation of noncrystalline solids through pressure-induced amorphization (PIA) processes. Amorphous materials produced via PIA are particularly interesting because, as disordered phases of originally perfectly ordered crystals, they offer a higher likelihood of being understood and engineered using well-established crystallographic theory. The probability of applying crystallographic principles from perfect crystals to interpret these amorphous structures is greater, given their origin from crystalline phases. Though amorphous structures can be realized through other routes \cite{doi:10.1080/00107516908204405,articleZhong} and PIA does not necessarily promote amorphization\cite{doi:10.1126/sciadv.abb3913}, it has been shown that internally consistent thermodynamic descriptions of this amorphization is possible \cite{1988Natur.334...52H}. Silica (SiO$_2$) is an archetypal material for the study of this phenomenon. The theory behind the reconstrucvtive phase transitions of SiO$_2$ polymorphs through pressurization is thoroughly explained by Dmitriev \cite{PhysRevB.58.11911}. Study of SiO$_2$'s amorphous forms is still of interest nowadays \cite{Zhang2020-bj}. Recent research has investigated PIA created amorphous solids as potential intermediates in a crystal-crystal phase transition \cite{Onodera2020-pl}. This introduces a fascinating avenue of study: formalizing an understanding of the topological long range order in crystals that potentially survives in their amorphous phases.

An important distinction in discussing order within amorphous structures lies between short, medium, and long-range order. Short-range order (SRO) refers to the local consistencies on the smallest scale, typically between individual atomic species \cite{doi:10.1038,He2024QuantifyingSO}. Medium-range order (MRO) extends beyond SRO, describing the structural organization of how local units connect and arrange themselves to fill three-dimensional space \cite{Sheng2006-aq}. In this work, we are particularly interested in the Long-range order (LRO) that represents the prototypical periodicity throughout the entire bulk of perfect crystals. Importantly, we aim to study the evolution of LRO in solids as a precursor to amorphous materials. An exaggerated example of this would be fullerene—carbon cages that exhibit LRO using amorphous building blocks \cite{Wang2012-sm}. Through pairwise distribution analysis, long-range ordering concepts, such as network topology, can be explored in glasses \cite{Salmon2005-dr}, offering intriguing research possibilities. Recent work has also examined hidden topological ordering in PIA materials, shedding light on the connections between amorphous disorder and crystal orientation, further advancing theoretical foundations for LRO in non-crystalline solids \cite{doi:10.1021/jacs.2c01717,10.1038/s41467-022-32419-5}. Nevertheless, significant research is still required across various materials and simulations to deepen our understanding of these phenomena.

Herein, we present the program \texttt{\texttt{PyLRO}}, an open source Python package built on top of Atomic Simulation Environment (\texttt{ASE}) \cite{ase-paper} to quantify and visualize the relative LRO of solids. Using structure files of moderately sized supercells of an amorphous material, without any required dependency on a reference parent structure, \texttt{\texttt{PyLRO}} can streamline in-depth analysis of long range periodicity in any arbitrary crystallographic direction. It provides intuitive quantification of order in each direction and displays the full projection of the structure's order on a crystallographic Miller sphere. We hope that this functionality can promote the study of topological order in amorphous materials by providing quick processing and visualization of computer-generated PIA structures during the amorphous-crystal transitions. In the following sections, we will start with introducing the basis behind the order quantification as well as the program's workflow. Next, we will demonstrate its application to a set of AlPO$_4$ data, a quartz prototype with a well-studied PIA transition. Finally, we conclude with a discussion of the program's potential uses in other areas of material research.

\section{Methodology}

\subsection{Long Range Order Quantification}

In this work, we aim to extract key structural information to quantify the long range order of a supercell structure in an unbiased and efficient manner. For a PIA, the input structure should not rely on a crystalline reference, as this would limit the program’s generality. Therefore, we seek to quantify LRO from only the atomic positions of a large supercell, without the use of reference crystal.


To avoid the unit cell complexity, a representative species needs to be selected to capture the repeating unit, and the long range order of the structure in each direction is assessed based on how closely the atomic spacings in each direction adhere to perfect periodicity. The task of quantifying the structure’s order is thus reduced to measuring the degree of atomic displacement from an idealized “perfect” crystal. Once this displacement is quantified, projections of both the structure and its idealized form onto any set of Miller planes can be used to illustrate the relative long-range order.

To achieve this goal, we develop \texttt{PyLRO} by starting with fitting a crystalline sub-lattice to the input amorphous structure. Specifically, the program seeks to obtain an optimal lattice beneath the atoms of the structure by minimizing their displacements from ideal positions. The calculation driving this fit is analogous to the structure factor used in simulating x-ray diffraction. The structure factor $S$ is defined as the squared modulus of the Fourier transform of the lattice, as expressed in equation \ref{Structure Factor}.

\begin{equation}
    S=\left|\sum^n_{i=1}e^{-2\pi i\mathbf{x_i}\cdot \mathbf{hkl}}\right|^2
    \label{Structure Factor}
\end{equation}

Traditionally, the structure factor is used to identify the directions in a crystal where scattering can produce non-zero intensity, making it a powerful tool for extracting periodicity characteristics. In our case, the atomic positions along the structure’s basis directions are projected onto a multi-period grid of plane waves. The period that yields the largest magnitudes from these projections is use as a determination of the optimal number of repeating units along the basis directions.

\begin{figure*}[t]

\begin{minipage}[c]{.58\textwidth}
\centering
\subfloat[]{\includegraphics[width=.95\textwidth]{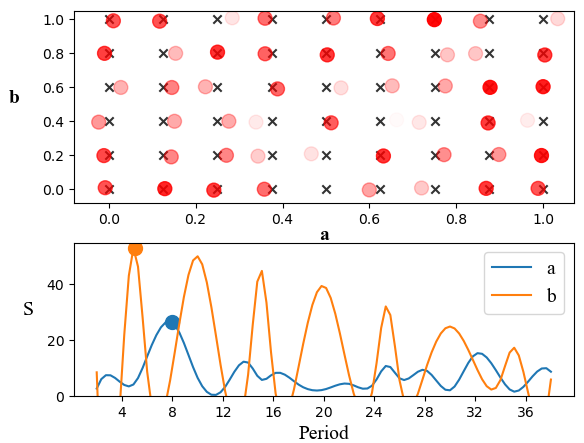} \label{multia}
}

\end{minipage}%
\begin{minipage}[c]{.4\textwidth}
\begin{center}
\scriptsize
\subfloat[]{
\begin{tabularx}{\linewidth}{>{\hsize=.04\hsize\centering}X >{\hsize=.25\hsize\centering}X >{\hsize=.25\hsize\centering}X >{\hsize=.17\hsize\centering}X >{\hsize=.21\hsize\centering\arraybackslash}X}

 & Atom Location & Sublattice Fit $\ell$  & Lattice Point $m$ & Disorder $d$\\
\hline
\hline
$n$  &  $(x,y)$  &  $(cx,dy)$   & $\text{round}(\ell)$ & $|m-\ell|$ \\
\hline
\hline
0  &  [-0.01, 0.01]  &  [-0.08, 0.05]  &  [0, 0]  &  [0.08, 0.05] \\
\hline
2  &  [0.24, 0.0]  &  [1.92, 0.0]  &  [2, 0]  &  [0.08, 0.0] \\
\hline
4  &  [0.46, 0.01]  &  [3.68, 0.05]  &  [4, 0]  &  [0.32, 0.05] \\
\hline
6  &  [0.72, 0.01]  &  [5.76, 0.05]  &  [6, 0]  &  [0.24, 0.05] \\
\hline
8  &  [0.99, 0.01]  &  [7.92, 0.05]  &  [8, 0]  &  [0.08, 0.05] \\
\hline
10  &  [0.14, 0.19]  &  [1.12, 0.95]  &  [1, 1]  &  [0.12, 0.05] \\
\hline
12  &  [0.34, 0.2]  &  [2.72, 1.0]  &  [3, 1]  &  [0.28, 0.0] \\
\hline
14  &  [0.63, 0.2]  &  [5.04, 1.0]  &  [5, 1]  &  [0.04, 0.0] \\
\hline
16  &  [0.9, 0.21]  &  [7.2, 1.05]  &  [7, 1]  &  [0.2, 0.05] \\
\hline
\multicolumn{5}{c}{\vdots}\\
\hline
38  &  [0.25, 0.81]  &  [2.0, 4.05]  &  [2, 4]  &  [0.0, 0.05] \\
\hline
40  &  [0.5, 0.79]  &  [4.0, 3.95]  &  [4, 4]  &  [0.0, 0.05] \\
\hline
42  &  [0.78, 0.79]  &  [6.24, 3.95]  &  [6, 4]  &  [0.24, 0.05] \\
\hline
44  &  [1.0, 0.79]  &  [8.0, 3.95]  &  [8, 4]  &  [0.0, 0.05] \\
\hline
46  &  [0.12, 0.99]  &  [0.96, 4.95]  &  [1, 5]  &  [0.04, 0.05] \\
\hline
48  &  [0.36, 1.01]  &  [2.88, 5.05]  &  [3, 5]  &  [0.12, 0.05] \\
\hline
50  &  [0.62, 1.01]  &  [4.96, 5.05]  &  [5, 5]  &  [0.04, 0.05] \\
\hline
52  &  [0.86, 0.99]  &  [6.88, 4.95]  &  [7, 5]  &  [0.12, 0.05] \\
\hline

\hline
\rule[-7pt]{0pt}{18pt}
 & \multicolumn{4}{r}{Average Disorder: $\overline{\textbf{d}}\textbf{=(0.14, 0.03)}$}\\
\end{tabularx}
\label{multib}
}

\end{center}
\end{minipage}
\captionsetup{width=.99\textwidth, justification=raggedright}
\caption[]{A toy model of an partially disordered structure with exaggerated disorder along the a axis. The left panel (a) demonstrates the evolution of structure factor $S$ as a function of period. 
The reference supercell dimension can thus be derived from $S$. The right panel (b) illustrates the procedures that quantify structural order based on the atomic deviations with respect to the reference unit cell.}  
\label{fig1}
\end{figure*}

In determining the supercell dimensions of the structure, cares must be taken to select a fit that maximizes the \textit{d}-spacing along the basis directions. For instance, a structure with three repeating units along a direction would produce similar scattering results from a wave with a 3 or 6 period. However, it would be incorrect to calculate disorder based on six repeating units, so the smallest possible multiple is always selected. Overall, we found that the confidence in \texttt{PyLRO}‘s ability to fit a crystalline sublattice to PIA structures is particularly high for those close to their crystalline phase. Nevertheless, this should not apply to study structures nearing complete disorder, where atomic displacements become so irregular that any quantified order would be negligible. Identifying the best possible crystalline sublattice is crucial, as disorder is primarily measured by the displacement of atoms relative to this calculated reference lattice.

After identifying an optimal sublattice reference, the next step is to calculate the precise displacements of each atom in the amorphous structure. By scaling the atomic positions to approximate their integer coordinates within the supercell, the displacements from these integer values can be understood as the fraction of a full period (or \textit{d}-spacing) that the atoms deviate in any given direction. To quantify long-range order along any Miller plane, we measure the average deviation of each atom from its ideal lattice point, expressed as a percentage of the lattice spacing in that specific direction. This allows for a consistent comparison of disorder across different directions, independent of the actual lattice spacing. Projections onto non-basis directions are always evaluated relative to the corresponding projection of the crystalline sublattice, which serves as the reference point for comparison.

Figure \ref{fig1} illustrates this design philosophy by quantifying the LRO of a simple two-dimensional toy model. In Figure \ref{multia}, the red points represent the atomic locations of the species selected to represent the repeating unit. The model demonstrates how significant amorphous deviations can cause atoms located near the boundary (due to periodic boundary conditions) to be displaced across to the opposite side of the supercell. Despite these minor deviations, the presence of long-range order in the structure is still evident and can be detected and quantified. Figure \ref{multia} also highlights how the calculated structure factor $S$ values for multiple periods help identify the lattice dimensions through periodic boundaries, without any prior knowledge of the reference crystal. Clear peaks are visible at 8 along the a-axis and at 5 along the b-axis. With these dimensions, \texttt{PyLRO} fits a lattice to the structure by (i) multiplying the atomic coordinates with the integer dimensions ($\ell$), (ii) detects the ideal sublattice values ($m$) by rounding the coordinates to the nearest integers; and (iii) compute the atomic disorder from the displacement between $\ell$ and $m$. Finally, whole structural disorder ($\overline{d}$) can be determined by taking the average of atomic disorder values ($d$). The entire procedure is also listed in the table of Figure \ref{multib}. 

\subsection{Miller Sphere Plot}
In addition to quantifying order, we are also interested in visualizing the overall long-range order of the entire amorphous structure graphically. To achieve this, \texttt{PyLRO} includes plotting functionality that generates a 3D Miller sphere — a unit sphere on a 3D plane defined by the $\mathbf{a}$, $\mathbf{b}$, and $\mathbf{c}$ basis vectors of the structure. Each point on the surface of the sphere corresponds to a specific Miller plane. Following the coauthor's recent work \cite{zhu2022quantification}, we use a series of Fibonacci points to provide an approximately equal distribution of points across the sphere’s surface, ensuring uniform sampling of directions.

For each point on the sphere, \texttt{PyLRO} calculates the long-range order parameter for the corresponding Miller plane and adds a Gaussian function scaled to 1 - $O_{hkl}$, where $O_{hkl}$ represents the calculated order parameter for the $\textit{hkl}$ plane. After normalization, the resulting figure warps the surface of the sphere to visually highlight the crystallographic directions with the highest long-range order. This visualization allows for a clear and intuitive understanding of the structure’s directional order, with the deformations of the sphere indicating regions of stronger or weaker order within the amorphous material.

An associated adjustable color map further enhances the visualization by clarifying the most ordered directions, with a contour overlay on the sphere’s surface to indicate regions of interest. Additionally, adjustable hyperparameters allow users to exaggerate or diminish the contrast between directions with strong and weak long-range order. These parameters can also control the sharpness of the peaks, making them more or less defined, depending on the desired level of detail. An example of this Miller sphere plot, including these features, will be discussed in the section \ref{Results}, demonstrating how the adjustable color map and hyperparameters help to distinguish and visualize the structure’s directional order more effectively.

\begin{figure}
    \centering
\includegraphics[width=.49\textwidth]{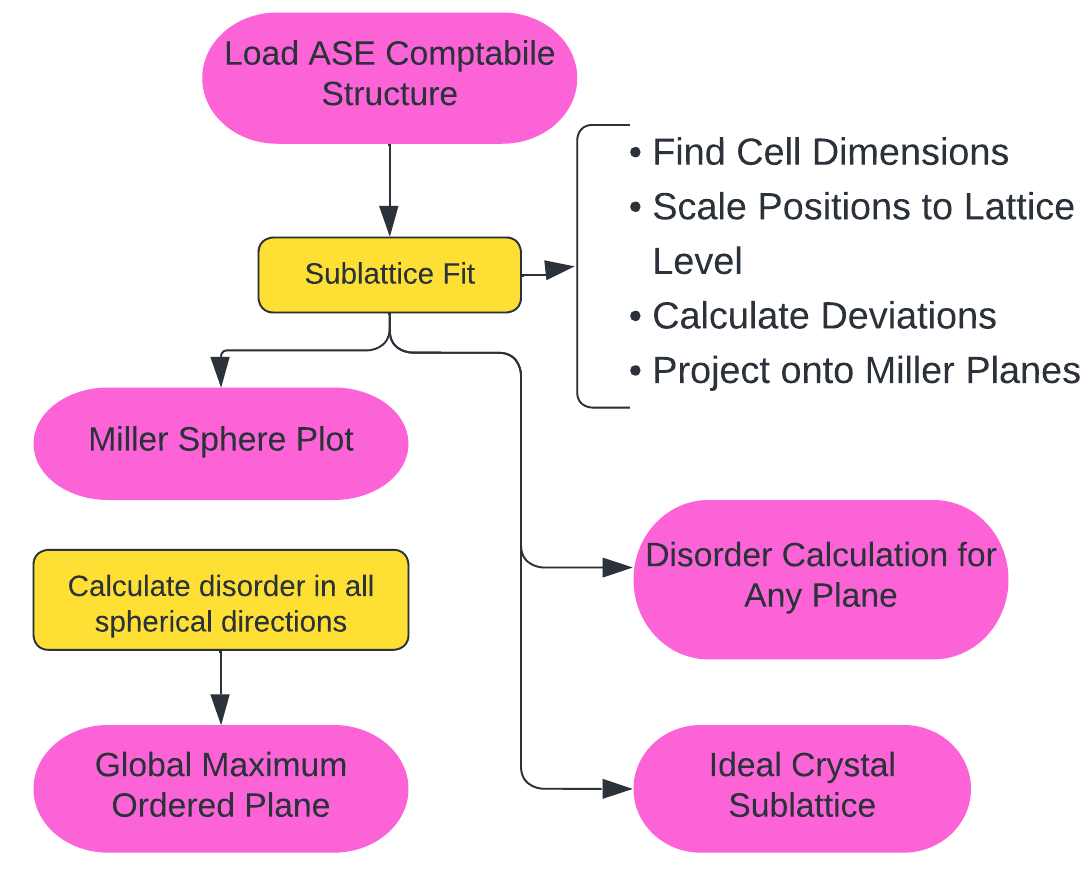}
    \captionsetup{justification=raggedright}
    \caption{An overview of \texttt{PyLRO} workflow.}
    \label{Workflow}
\end{figure}
\begin{figure*}[t]

\begin{minipage}[c]{.57\textwidth}
\centering
\subfloat[]{\includegraphics[width=.95\textwidth]{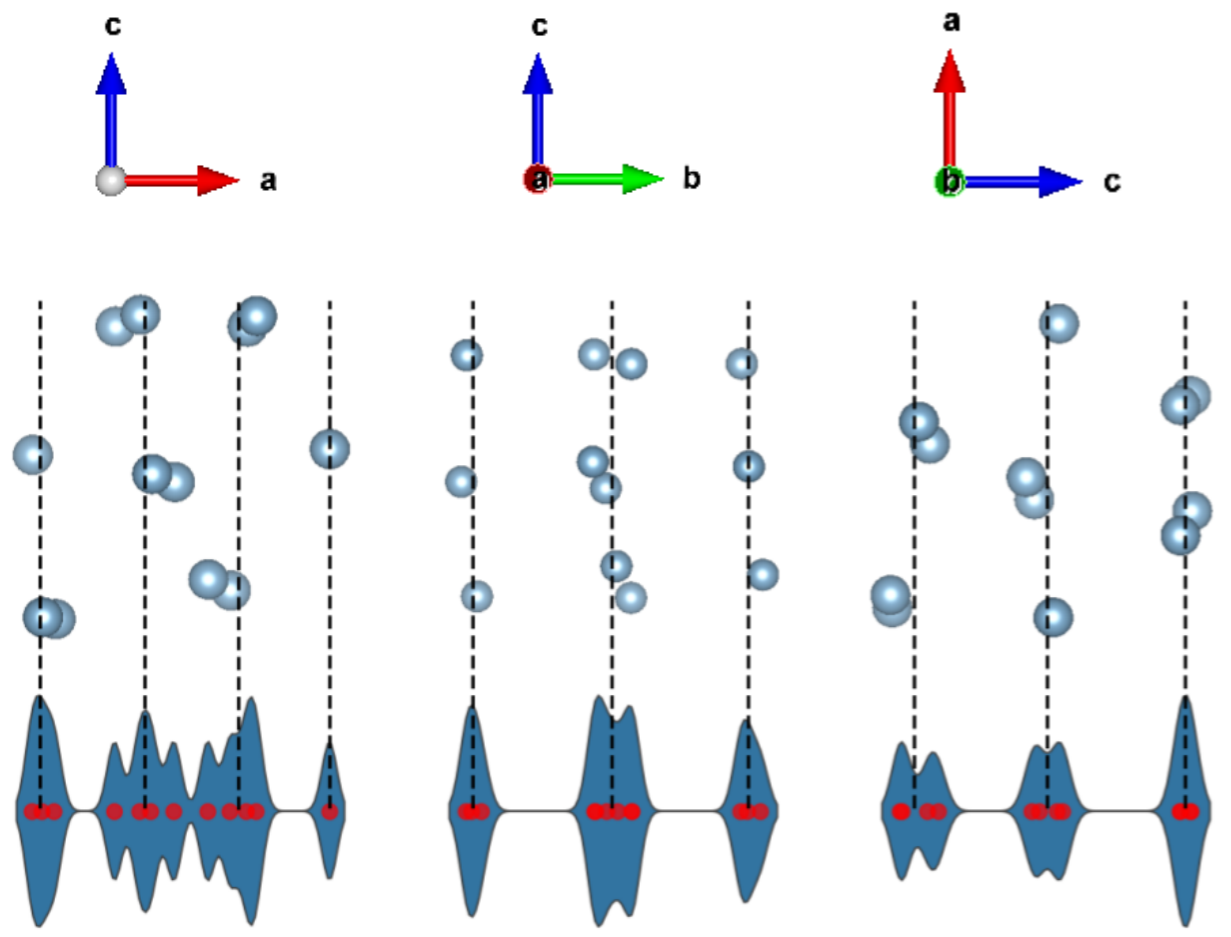} \label{AlPO4 multiview}
}

\end{minipage}%
\begin{minipage}[c]{.43\textwidth}
\scriptsize
\subfloat[]{\includegraphics[width=.94\textwidth]{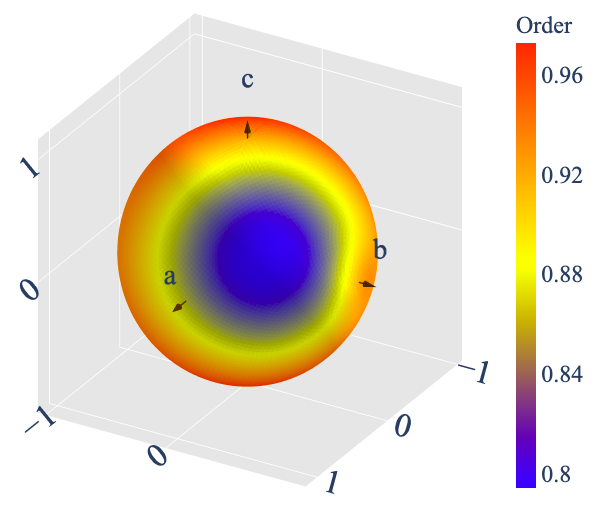}
\label{AlPO4 Miller Sphere}}

\end{minipage}
\captionsetup{width=.99\textwidth, justification=raggedright}
\caption[]{(a) Aluminum sublattice in the sample AlPO$_4$ structure visualized by the \texttt{VESTA} software \cite{vesta}  in three basis directions. The violin plots demonstrate the degree of deviation of the atoms from their respective sublattice positions. The calculated disorders are 0.145  along \textbf{a}, 0.112 along \textbf{b}, and 0.084  along \textbf{c}. (b) the associated Miller sphere plot for all directions, suggesting that of, the basis direction, \textbf{c} exhibits the maximum order while a direction close to the \textbf{ab} plane ([0.665, 0.602, 0.441]) possesses the minimum order.}
\label{new_multii}
\end{figure*}

\subsection{Workflow}
The final demonstration of the program’s workflow is shown in Figure \ref{Workflow}. In short, the process begins by loading an ASE-compatible structure file \cite{ase-paper}, which represents a supercell of a partially disordered structure. The program then automatically detects the optimal sublattice beneath the structure using the theory described earlier. Once the sublattice is established, calculations of the average disorder in any given plane, or the identification of the most (or least) disordered global plane, can be performed. Finally, the program generates a visual diagram of the structure’s global order through a Miller sphere plot, offering an intuitive way to visualize the directional order.

The core of the order quantification logic focuses on identifying the ideal sublattice for the amorphous structure. Once this reference is established, fundamental principles of crystal periodicity are applied to measure atomic displacements and deviations in any direction. The plotting functionality leverages the complete quantification of disorder across all planes, enabling users to easily compare the relative long-range order between different directions. This workflow not only automates the key steps in quantifying structural order but also provides clear visual representations to facilitate deeper insights into the degree of order and disorder within amorphous systems.

\section{Results and Discussions}
\label{Results}

AlPO$_4$ is a archetypal memory glass material for this kind of study. It has had many of its properties already thoroughly outlined in the literature, such as its vibrational modes \cite{PhysRevB.35.8316}, pressure induced transitions \cite{10.2138/am.2013.3897, 10.1038/nmat1966}, and potential memory glass effects\cite{PhysRevB.51.11262, PhysRevLett.71.3143, michael_b__kruger__1990}. In a recent analysis into topological ordering of memory glass on extended time scales \cite{doi:10.1021/jacs.2c01717}, an exploration of the AlPO$_4$ energy landscape under pressure found a \textbf{S}warm of highly probable \textbf{L}ow \textbf{S}ymmetry (SLS) structures connecting the high pressure phase of berlinite to the global minimum AlPO$_4$ II. In the following section, we analyze the structures in this SLS using \texttt{PyLRO} to validate the function's accuracy against the current accepted theory around AlPO$_4$'s long range topological order.

\subsection{Case Study on a Sample Structure}
We begin with a sample structure. Figure \ref{AlPO4 multiview} shows the aluminum substructure from three views. The projections onto the \textbf{ab} and \textbf{bc} planes show through visual inspection the degree to which the Al atoms deviate from the ideal crystal. While there is not much to choose from between the \textbf{a} and \textbf{b} directions, it is clearly seen that the \textbf{c} direction contains quite significant order. This is consistent with the calculated order parameters in each crystallographic basis direction, where \textbf{c} has the lowest disorder at 0.084. The average percent deviation of the atoms in the \textbf{c} direction is just 8.4\% of the [001] \textit{d} spacing. 

\begin{figure*}[t]

\begin{minipage}[c]{.45\textwidth}
\centering
\subfloat[]{\includegraphics[width=0.99\linewidth]{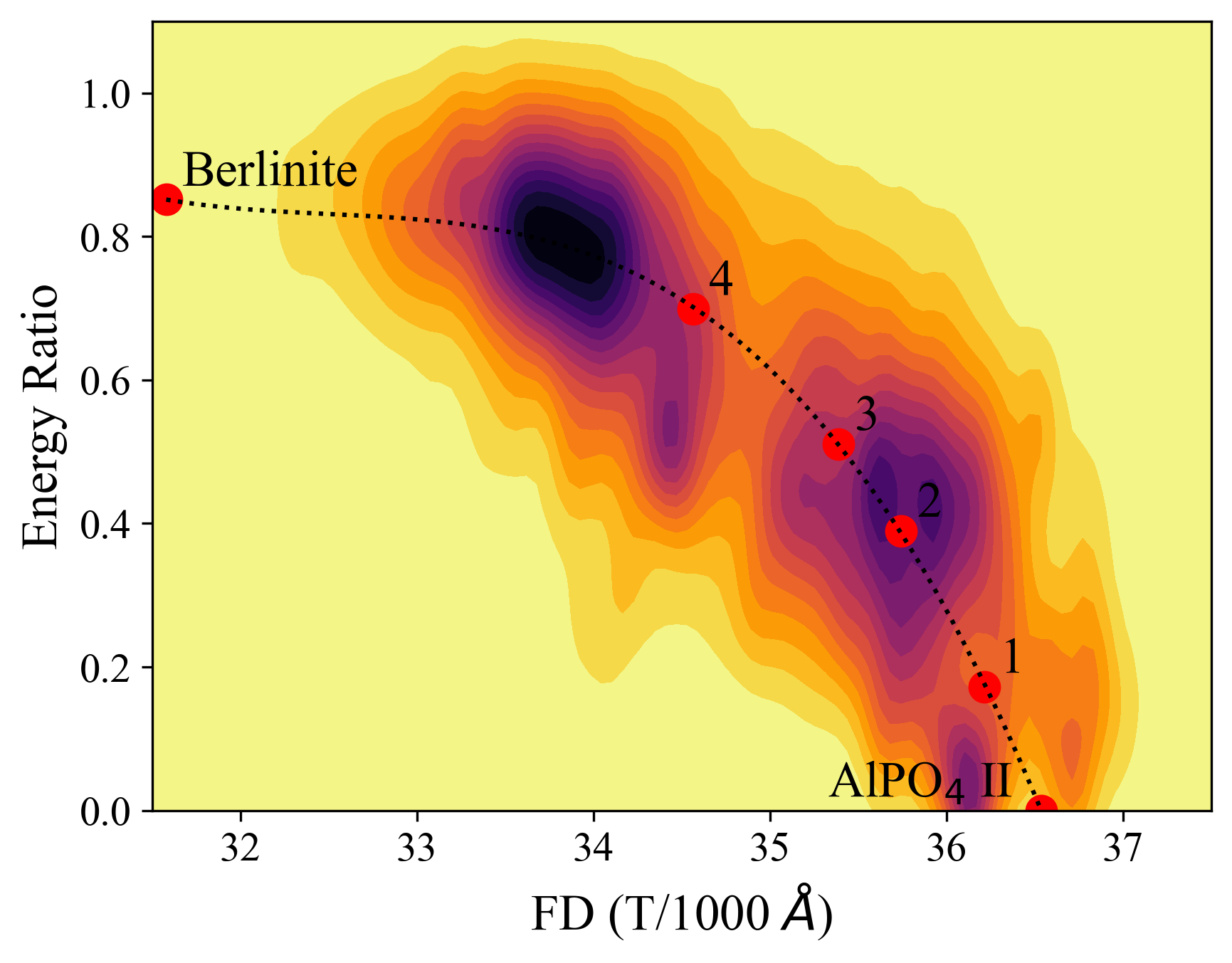}\label{AlPO4 Energy Landscape}}
\end{minipage}
\begin{minipage}[c]{.5\textwidth}
\centering
\subfloat[]{
\begin{tabular}{ccc}
\includegraphics[width=.3\textwidth]{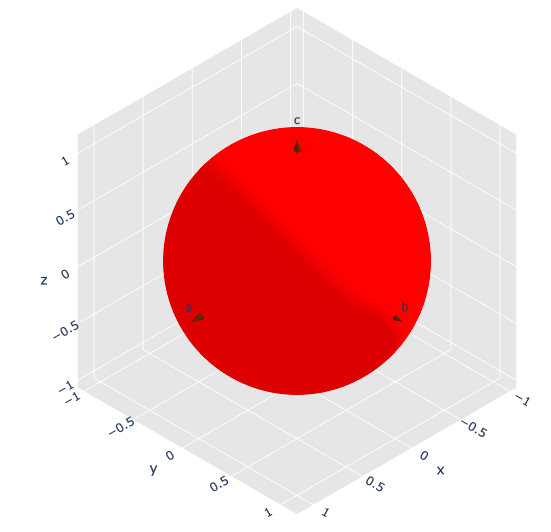} & \includegraphics[width=.3\textwidth]{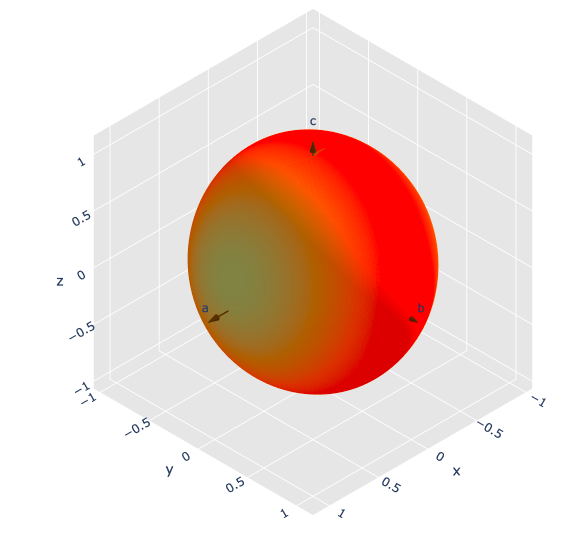} & \includegraphics[width=.3\textwidth]{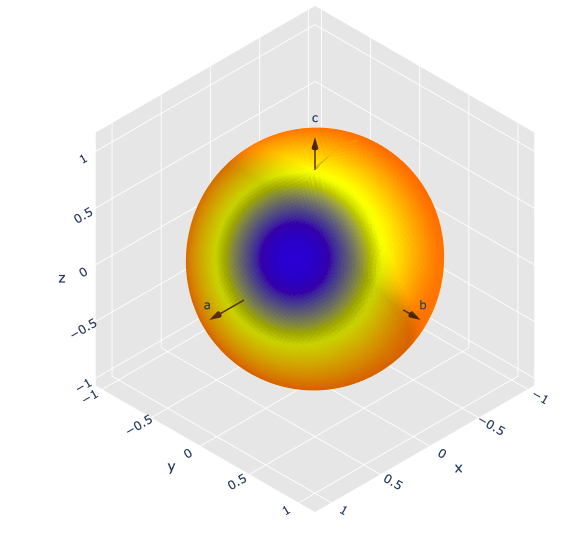} \\
(AlPO$_4$ II) 1.000 & (1) 0.057 & (2) 0.248\\

\includegraphics[width=.3\textwidth]{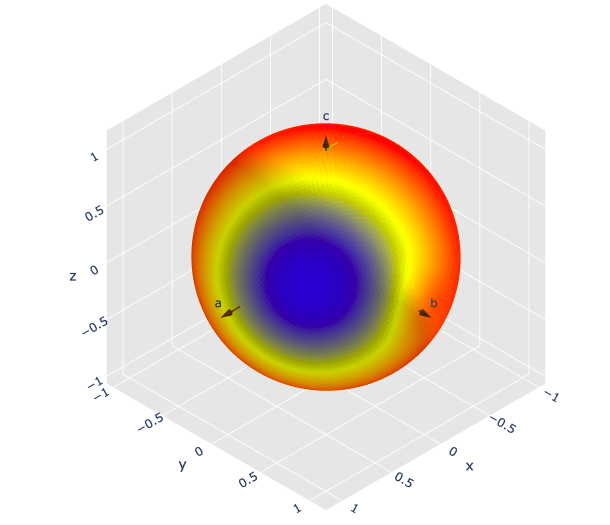} & \includegraphics[width=.3\textwidth]{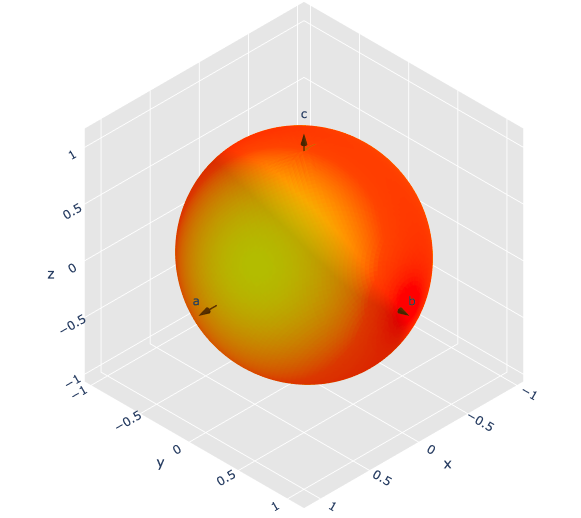} & \includegraphics[width=.3\textwidth]{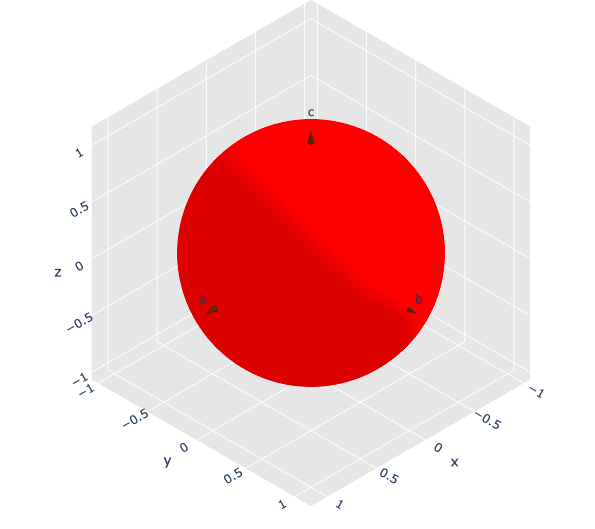} \\
(3) 0.110 & (4) 0.059  & (Berlinite) 1.000\\
\end{tabular}
\label{b}}
\end{minipage}
\captionsetup{width=.99\textwidth, justification=raggedright}
\caption{(a) An energy landscape of the AlPO$_4$ SLS at 15 GPa against the structural framework density (FD). The high pressure phase of berlinite (labelled as 6) is toward the upper left while the global minimum AlPO$_4$ II (labelled as 1) is at the bottom. 
(b) list the evolution of Miller sphere plots of six representative structures along the transition trajectory from AlPO$_4$ II to berlinite. The values given is the average disorder among the basis directions.}
\label{bigfigure}

\end{figure*}

Figure \ref{AlPO4 Miller Sphere} captures this information and further analyzes a complete description of the structure's order in all directions. It shows that most ordered directions globally are a ring of planes around the miller surface with the \textbf{c} being most ordered basis direction. These directions in real space can be found through their products with the structure's basis matrix. The ring-like ellipsoid makes theoretical sense. The highest ordered directions should be perpendicular to the lowest order direction as the directions of maximum atomic deviations have been projected out. We would also expect a smooth plot because the contribution of any directional disorder should change continuously as the analysis window moves across directions where it is sampled more or less.

These results agree with and formalizes further the community acceptance of AlPO$_4$'s amorphous transition pathways. It was found by Zhu and coworkers \cite{doi:10.1021/jacs.2c01717} that the ideal description to understand amorphous AlPO$_4$ is that of an extreme Carpenter-type crystalline transition \cite{Carpenter1998-ga} as its signature memory glass effects rely on the long range order that survives as crystal AlPO$_4$ II deforms topologically, the relatively low energy barrier heights of which can be  can  through the integral of spontaneous strains from Carpenter's theory\cite{carpenter1998}. Transition pathway analysis showed that while the medium range order describing the relative orientation of the characteristic Al-/P- polyhedra was broken, the short range order of the polyhedra structures was kept mostly intact and the long range order is partially kept. Figure \ref{AlPO4 Miller Sphere} shows that the \textbf{c} direction of the chosen representative structure of the SLS retains a healthy amount of long range order, preserving the connectivity of the structure's building blocks along that direction. Further, \textbf{a} and \textbf{b} being disordered in relatively equal amounts supports the current understanding of AlPO$_4$'s spiral ring symmetries during amorphization: the long range  symmetry in the \textbf{c} direction remains topologically strong as the symmetries within the rings in the \textbf{ab} plane break down. We can analyze these results through the lens of a ringed spiral structure for additional insight. The ``backbone" of such a structure should be the even spacing of polyhedral rings along the spiral \textbf{c}. For it to deform topologically, it makes sense the rings should not deviate far in this direction as losing this particular long range order would mean polyhedra are no longer structured as a chain, making it unlikely to retain its crystalline ``memory". The results verify that this long range order of this disordered structure is strong relative to the ring plane, agreeing with this logic.

\begin{figure*}
\centering
\includegraphics[width=.99\linewidth]{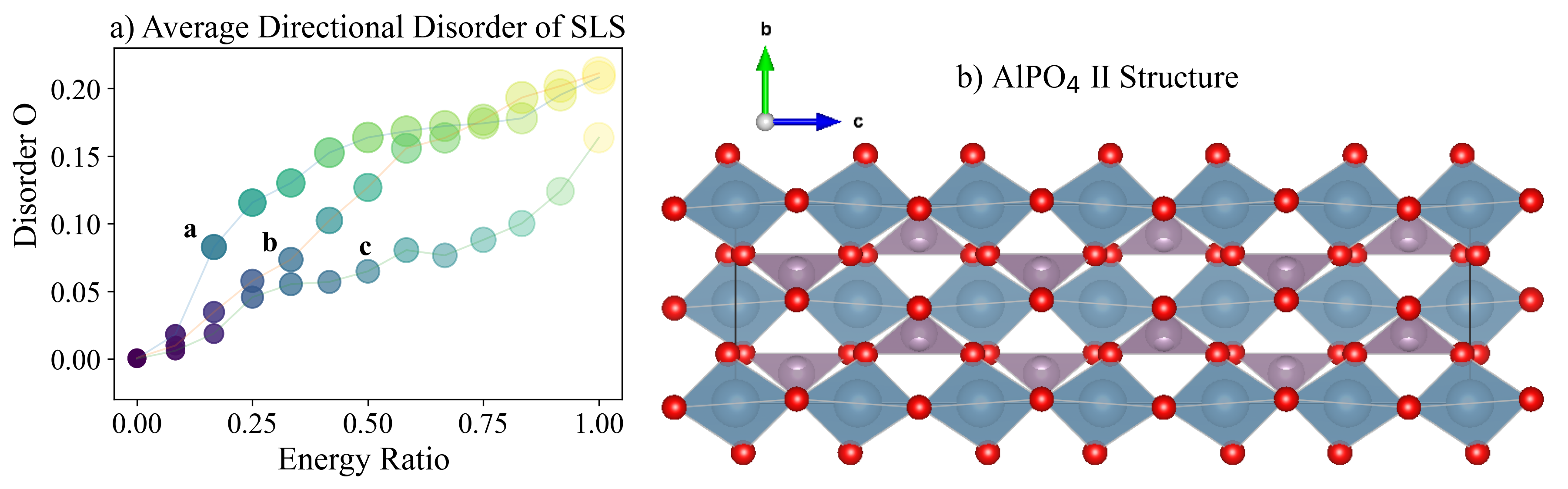}
\includegraphics[width=.99\linewidth]{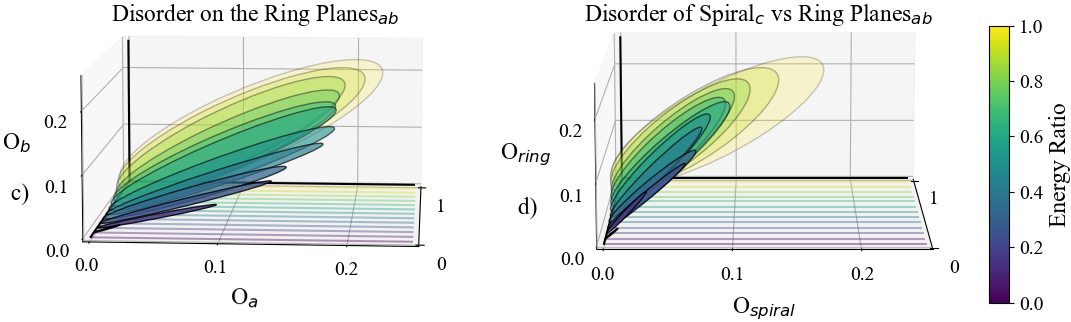}
\captionsetup{width=.99\textwidth, justification=raggedright}
\caption{(a) Plots the propagation of the averaged PyLRO calculated directional disorder of structures in regions of increasing energy in the SLS; (b) displays the crystalline ALPO$_4$'s spiraling polyhedra along the \textbf{c} direction; (c) displays a waterfall plots showing the evolution of disorder in the structures of the same regions between the directions of the ring plane \textbf{a} vs \textbf{b}; (d) is another waterfall plot showing showing the evolution of disorderbetween the spiral direction \textbf{c} vs planar disorder \textbf{ab}.}
\label{endfigure}
\end{figure*}

\subsection{High-throughput Analysis on a Massive Dataset}

One of the primary benefits of \texttt{PyLRO} is its lightweight, allowing for high throughput calculations on structure sets that would otherwise be untenable. Figure \ref{AlPO4 Energy Landscape} shows the density of states for the SLS created using 9533 structures from the coauthor's previous study \cite{doi:10.1021/jacs.2c01717}. Knowing the global minimum AlPO$_4$ II, we can do disorder analysis on amorphous structure in the higher energy ranges to identify patterns in the topological order as symmetries get broken. Figure \ref{b} shows Miller sphere plots of structures along a trajectory connecting AlPO$_4$ II to Berlinite. It can be seen that the AlPO$_4$ initially shows perfect symmetry. As we move up the energy landscape, we can see that disorder does not affect the structures randomly. The structures begin to lose translational symmetry along the \textbf{a} direction first. The disorder manifests more so in the \textbf{b} direction further, and the order in the \textbf{c} direction throughout the entire trajectory remains strong and clearly better than that of \textbf{a} and \textbf{b}.

Furthermore, Figure \ref{endfigure} provides a broader analysis of the average directional disorder across the entire energy landscape. The average disorder in the $\mathbf{a}$, $\mathbf{b}$, and $\mathbf{c}$ directions is calculated for groups of structures within preceding 1 eV increments. All energy values are represented as a ratio of the energy range between AlPO$_4$ II and berlinite. 
The results clearly show that the average disorder increases in all directions as we move higher up the energy landscape. Notably, the $\mathbf{a}$ direction remains more disordered than the other basis directions up to approximately 0.8 eV, at which point the disorder in the $\mathbf{b}$ direction begins to match it. This observation is consistent with the Miller sphere plots in Figure \ref{b}. As expected, the $\mathbf{c}$ direction remains the most ordered on average throughout the energy landscape, suggesting that long-range topological order is preserved within the amorphous structures.

Using \texttt{PyLRO}, we can also conveniently visualize the anisotropic amorphization process in the two dimensional plane. Figures \ref{endfigure}c and \ref{endfigure}d present waterfall ellipse plots showing the evolution of disorder direction and magnitude within the ring plane of polyhedra, as well as the disorder along the topologically strong $\mathbf{c}$ direction relative to the average disorder in the ring plane. \texttt{PyLRO}’s analysis across the energy landscape highlights how the topology of the structures changes. In the ring plane, the disorder in the $\mathbf{a}$ direction initially decreases the most. However, as disorder increases, it equalizes with that along the $\mathbf{b}$ direction. Interestingly, the disorder in the spiral $\mathbf{c}$ direction consistently remains lower than that of the $\mathbf{ab}$ ring plane, even as the overall disorder in the structures continues to increase.

Finally, the analysis reveals that independently calculated measures of the directions and magnitudes of maximum and minimum disorder (corresponding to the major and minor axes of the ellipse) are consistently perpendicular, as expected from symmetry considerations. This consistency further validates the accuracy of \texttt{PyLRO} in capturing both the magnitude and directional aspects of disorder within amorphous systems.

\section{Conclusion}
The capabilities of the open-source Python package \texttt{PyLRO}, particularly its ability to calculate degrees of directional disorder, have been thoroughly demonstrated in this work. We provided detailed explanations of the program’s workflow and design logic, illustrated through a simple amorphous model. For a rigorous test, we showcased \texttt{PyLRO}’s performance using the well-studied material AlPO$_4$. By leveraging a large dataset of low-symmetry AlPO$_4$ structural intermediates, we demonstrated how \texttt{PyLRO} efficiently calculates and visualizes the degree of directional disorder for both individual structures and entire structure sets. Our analysis of a single sample revealed the long-range topological order that has been proposed as the underlying mechanism behind the structural memory effects in amorphous AlPO$_4$. Moreover, we showed how this long-range order can be rigorously quantified, matched with crystallographic intuition, and compared across different amorphous structures to provide a meaningful assessment of order and disorder.

Using \texttt{PyLRO}, we were able to perform a high-throughput analysis of a vast dataset comprising 9,533 structures. This allowed us to gain insight into the pressure-induced amorphization of AlPO$_4$. Its utility extends to time-series data generated from molecular dynamics simulations or other advanced computational sampling methods. We believe that \texttt{PyLRO} will prove valuable to a wide range of researchers who are interested in studying the amorphous-crystal transition.

\section*{Acknowledgments}
This research is primarily sponsored by the U.S. Department of Energy, Office of Science, Office of Basic Energy Sciences and the Established Program to Stimulate Competitive Research (EPSCoR) under the DOE Early Career Award No. DE-SC0021970. QH was supported by the National Natural Science Foundation of China with grant No. T2425016 and U2230401. The computing resource is provided by ACCESS (TG-DMR180040).

\section*{Data availability}
The \texttt{PyLRO} source code, instructions, as well as scripts used to calculate the results of this study, are available in \url{https://github.com/MaterSim/PyLRO}.

\bibliography{ref.bib}

\end{document}